\documentclass{article} % For LaTeX2e

\usepackage{iclr2026_conference,times}

\usepackage{amsmath,amsfonts,bm}

\def\eqref#1{equation~\ref{#1}}
\def\1{\bm{1}}

\DeclareMathAlphabet{\mathsfit}{\encodingdefault}{\sfdefault}{m}{sl}
\SetMathAlphabet{\mathsfit}{bold}{\encodingdefault}{\sfdefault}{bx}{n}

\usepackage{hyperref}
\usepackage{url}
\usepackage{graphicx}
\graphicspath{{./figure/}{./table/}}

\usepackage{amsmath}
\usepackage{amssymb}
\usepackage{amsfonts}

\usepackage{booktabs}
\usepackage{multirow}
\usepackage{float}
\usepackage{pdflscape}

\title{Portfolio Optimization under Recursive Utility via Reinforcement Learning}

\author{Minkey Chang}

\iclrfinalcopy % Camera-ready/non-anonymous version.

\begin{document}

\maketitle

\begin{abstract}
We study whether a risk-sensitive objective from asset-pricing theory---recursive utility---improves reinforcement learning for portfolio allocation. The Bellman equation under recursive utility involves a certainty equivalent (CE) of future value that has no closed form under observed returns; we approximate it by $K$-sample Monte Carlo and train actor-critic (PPO, A2C) on the resulting value target and an approximate advantage estimate (AAE) that generalizes the Bellman residual to multi-step with state-dependent weights. This formulation applies only to critic-based algorithms. On 10 chronological train/test splits of South Korean ETF data, the recursive-utility agent improves on the discounted (naive) baseline in Sharpe ratio, max drawdown, and cumulative return. Derivations, world model and metrics, and full result tables are in the appendices.
\end{abstract}

% Main sections
\section{Introduction}

Modern portfolio agents built with reinforcement learning \cite{sutton2018reinforcement} focus largely on algorithms such as PPO and actor-critic (A2C) \cite{mnih2016asynchronous,schulman2017proximal}, with much of the literature aimed at algorithmic improvements: more efficient optimization, scalable or multi-agent setups, and better exploration. We argue that economic theory offers a complementary direction. Utility-based frameworks assume coherent, rational preferences over risk and time \cite{epstein1989substitution,campbell2002strategic} and are a natural fit for machine agents that do not suffer the biases of human decision-makers. In this paper we ask whether equipping conventional RL with a different \emph{objective}---recursive utility from asset-pricing theory \cite{epstein1989substitution}---improves portfolio outcomes.

Recursive utility separates risk aversion from intertemporal substitution. We integrate it into actor-critic RL via a sampling-based certainty equivalent and use the resulting value target and advantage for policy updates; it applies only with critic-based methods (PPO, A2C). On 10 train/test splits, the recursive-utility agent improves on the naive baseline in Sharpe ratio and max drawdown.

\section{Related Work}

\paragraph{Portfolio theory.}
Mean--variance optimization \cite{markowitz1952,sharpe1964capital} and consumption--portfolio rules \cite{merton1971optimum} with long-horizon theory \cite{campbell2002strategic} form the classical foundation; equal-weight (1/N) rebalancing remains a strong benchmark. Recursive utility \cite{epstein1989substitution,backus2004asset} separates risk aversion from intertemporal substitution and yields hedging demand; portfolio choice under recursive preferences is studied in \cite{liu2007portfolio} and \cite{schroder1999optimal}; \cite{campbell2002strategic} derive approximate rules under predictable returns, which we use as policy priors.

\paragraph{Reinforcement learning and portfolios.}
RL agents maximize cumulative reward via policy gradient or actor-critic methods \cite{sutton2018reinforcement,graesser2020foundations}. Actor-critic and A2C \cite{mnih2016asynchronous} and PPO \cite{schulman2017proximal} are standard; deep RL for portfolio management typically uses additive rewards and linear Bellman targets \cite{jiang2017deep,liu2020finrl}, with some work on risk-sensitive or adaptive objectives \cite{almahdi2017adaptive}. \cite{dixon2023time} treat consumption under recursive utility in RL. We put recursive utility \emph{directly} in the value target and advantage and restrict it to critic-based algorithms (PPO, A2C). World model, environment, and metrics are in Appendix~\ref{app:environment}.

\section{Method}
\label{sec:method}

We use recursive utility in an RL framework. The Bellman equation involves the certainty equivalent (CE) of future utility; with no closed form under observed returns, we approximate the CE by $K$-sample Monte Carlo and train actor-critic on the resulting value target. Recursive utility is only defined for critic-based algorithms (e.g., PPO, A2C), because the CE and the Bellman-residual advantage both require a learned value function $V_\phi$.

\subsection{Problem Setup}

We treat portfolio allocation as a partially observable Markov decision process: the agent observes asset returns and its own wealth and positions, but not the full market state. Time is discrete; at each step $t$ the agent chooses a portfolio and wealth evolves according to the realized returns.

\paragraph{Wealth and state.}
Let $W_t > 0$ denote \emph{wealth} (dollar value of the portfolio). We work with \emph{log wealth} $w_t = \log W_t$ in the state and in the Bellman equation. Log wealth is used for numerical stability and scale-invariance: returns are additive in $w_t$ ($w_{t+1} - w_t = \log(1 + R_{\mathrm{port},t+1})$), and the recursion does not depend on the absolute level of wealth when preferences are homogeneous. We normalize initial wealth so that $W_0 = 1$ and $w_0 = 0$. The state at time $t$ is $s_t = (w_t, \boldsymbol{\alpha}_{t-1})$, where $\boldsymbol{\alpha}_{t-1} \in \mathbb{R}^n$ are the portfolio weights chosen at the previous step (fractions of wealth in each of $n$ assets).

\paragraph{Wealth dynamics.}
Given portfolio weights $\boldsymbol{\alpha}_t$ and realized asset returns $\mathbf{R}_{t+1}$, the portfolio gross return is $1 + \boldsymbol{\alpha}_t^\top \mathbf{R}_{t+1}$, so wealth evolves as $W_{t+1} = W_t (1 + \boldsymbol{\alpha}_t^\top \mathbf{R}_{t+1})$. In log space,
\begin{equation}
w_{t+1} = w_t + \log(1 + \boldsymbol{\alpha}_t^\top \mathbf{R}_{t+1}).
\end{equation}

\paragraph{Actions and constraints.}
Portfolio weights must lie in the simplex $\Delta^n$ (non-negative and sum to one). In our implementation the policy outputs \emph{increments} $\Delta \boldsymbol{\alpha}_t$ that are added to the previous weights and then projected onto $\Delta^n$; this incremental parameterization improves learning stability compared to predicting raw weights directly. The projection and the optional direct (raw-weight) parameterization are described in Appendix~\ref{app:environment}. We also support an optional three-layer architecture (Appendix~\ref{app:three_layer}). The objective is recursive utility; the Bellman equation and its derivation are in Appendix~\ref{app:ez_derivation}.

\subsection{Recursive Utility and the Role of Consumption}

In the standard economic formulation, recursive utility is defined over \emph{consumption}: current consumption yields immediate utility, and the trade-off between consuming today and leaving wealth for tomorrow drives consumption smoothing and a more precautionary attitude toward the future. For a human, consumption is the flow of spending that generates well-being; it is also what we ``give up'' when we save, which shapes how we perceive and weight future outcomes. For an \emph{artificial agent}---a trading system or portfolio manager---there is no literal consumption: the machine does not consume in the sense that humans do. Our implementation is motivated by the view that consumption in the aggregator is not merely a formal device to retain the recursive structure; we interpret it as the \emph{source} of strategic, forward-looking decision-making. The trade-off between ``consuming'' today and preserving value for tomorrow can be thought of as a survival instinct: the agent behaves as if something is at stake (current versus future payoff), which induces more careful, risk-aware choices. Retaining this trade-off in the objective is therefore a design choice for better decision-making, not only for mathematical consistency.

We introduce consumption as an \emph{accounting} variable: we let current ``consumption'' be parameterized by a scalar $\kappa_t$ (kappa) that scales with wealth, e.g.\ $C_t = \kappa_t W_t$. The agent does not literally consume; $\kappa_t$ is a free parameter (learned or fixed) that plays the role of the consumption-to-wealth ratio in the utility aggregator. This preserves the intertemporal trade-off that we interpret as the driver of survival-oriented behavior, while remaining implementable for a non-consuming agent. In effect, $\kappa$ controls how much weight is placed on ``current-period'' utility versus the certainty-equivalent of future utility. The full Bellman equation and the definition of the certainty equivalent are given in Appendix~\ref{app:ez_derivation}; here we state only the form of the value target used in learning.

\paragraph{Value target and certainty equivalent.}
Recursive utility \cite{epstein1989substitution} aggregates a current-period term (involving $C_t = \kappa_t \exp(w_t)$) and the \emph{certainty equivalent} (CE) of next-period utility. Risk aversion $\gamma$ enters through the CE: higher $\gamma$ penalizes downside dispersion of future value. The Bellman equation is
\begin{equation}
V(s_t) = \max_{\boldsymbol{\alpha}_t, \kappa_t} \Big\{ (1-\beta)(\kappa_t \exp(w_t))^{1-1/\psi} + \beta \Big(\mathbb{E}\big[V(s_{t+1})^{1-\gamma} \mid s_t, \boldsymbol{\alpha}_t, \kappa_t\big]\Big)^{\frac{1-1/\psi}{1-\gamma}} \Big\}^{1/(1-1/\psi)}.
\label{eq:bellman}
\end{equation}
We do not have a closed form for the expectation under the observed return distribution, so the CE must be approximated; see the next subsection and Appendix~\ref{app:ez_derivation}.

\subsection{Estimation}

We need to estimate the certainty equivalent (CE) of future utility, the value target, and the advantage. The CE aggregates the distribution of next-period value into a single summary that reflects risk aversion; when $\gamma > 1$ it penalizes downside dispersion, which is what distinguishes recursive utility from a simple discounted sum of rewards. The Bellman equation \eqref{eq:bellman} is \emph{non-linear} in the expectation---the CE is a power mean of $V(s_{t+1})^{1-\gamma}$, not a linear expectation---and under the observed return process we do not assume a tractable closed form for the distribution of $s_{t+1}$. The expectation $\mathbb{E}[V(s_{t+1})^{1-\gamma} \mid s_t, \boldsymbol{\alpha}_t, \kappa_t]$ is therefore not feasible to compute exactly. We estimate it by \emph{sampling}: we draw $K$ next-state candidates (by sampling returns from an empirical or model-based distribution), evaluate the critic on each, and form the empirical power mean. With $s_{t+1}^{(1)}, \ldots, s_{t+1}^{(K)}$ the $K$ sampled next states,
\begin{equation}
\widehat{\text{CE}}_t^{(K)} = \Big( \frac{1}{K} \sum_{k=1}^{K} V_\phi\big(s_{t+1}^{(k)}\big)^{1-\gamma} \Big)^{1/(1-\gamma)}.
\label{eq:ce_sample}
\end{equation}
This estimator is consistent: as $K \to \infty$ it converges to the true CE when the samples are drawn from the conditional distribution of $s_{t+1}$. The recursive utility Bellman operator is a contraction in the certainty-equivalent metric \cite{epstein1989substitution,schroder1999optimal}; see Appendix~\ref{app:ez_derivation}. Replacing the exact CE with a consistent estimator preserves the fixed-point structure and justifies using $\widehat{\text{CE}}_t^{(K)}$ in the value target. We estimate the value target as
\begin{equation}
\hat{T}_t^{EZ} = \Big( (1-\beta)(\kappa_t \exp(w_t))^\rho + \beta \big(\widehat{\text{CE}}_t^{(K)}\big)^\rho \Big)^{1/\rho}.
\label{eq:ez_target}
\end{equation}
Return distribution for sampling and batched env steps are in Appendix~\ref{app:environment}; hyperparameters are in Appendix~\ref{app:implementation}. We train the critic to match $\hat{T}_t^{EZ}$ with MSE and estimate the advantage from the Bellman residual (and optionally a multi-step extension; Appendix~\ref{app:aae}):
\begin{align}
L_{\text{critic}} &= \mathbb{E}\big[\big(V_\phi(s_t) - \hat{T}_t^{EZ}\big)^2\big], \label{eq:critic_loss} \\
A_t^{EZ} &= \hat{T}_t^{EZ} - V_\phi(s_t). \label{eq:ez_advantage}
\end{align}
The policy is updated with PPO \cite{schulman2017proximal} using $A_t^{EZ}$; Campbell--Viceira rules \cite{campbell2002strategic} initialize the actor (Appendix~\ref{app:ez_derivation}).

\section{Experiments}
\label{sec:experiments}

\subsection{Dataset}

We use daily ETF closing prices from the South Korean stock market. The universe consists of 110 ETFs selected by listing on the Korean market, sufficient history over the sample period, and exclusion of names with excessive missing data; criteria and preprocessing are in Appendix~\ref{app:environment} (World Model and Environment). The data (\texttt{data/data.csv}) are split into 10 chronological train/test splits (train ratio 50\%--90\%, split1--split10). Each split is used for training and evaluation separately; we report mean $\pm$ std over the 10 splits \cite{liu2020finrl}.

\subsection{Environment and Algorithm}

The environment is the portfolio MDP described in Section~\ref{sec:method}: state (log wealth, previous weights), simplex-constrained actions, and wealth dynamics from realized returns. We compare three objectives: naive (discounted sum of portfolio returns), Markowitz (per-step mean--variance reward), and recursive utility (CE-based value target and advantage). Algorithms are Random, REINFORCE, A2C, and PPO. Recursive utility is supported only with critic-based methods (PPO, A2C), consistent with the requirement for a value function in the CE and advantage. Environment and data are identical across objectives; only the Bellman target and advantage differ for recursive. We evaluate on the test set using Sharpe ratio (SR), Sortino, Calmar, max drawdown (MDD\%), cumulative return (CR\%), and volatility (Vol\%). One trial per split; metric definitions are in Appendix~\ref{app:environment} (Metrics).

\subsection{Results}

With $\gamma > 1$, the CE penalizes downside dispersion (Section~\ref{sec:method}), so we expect a more risk-aware policy. Recursive attains higher SR (2.07 vs 1.22) and lower MDD (10.38\% vs 12.26\%) than naive, and higher average CR (8.23\% vs $-$6.47\%); the improvement in both risk-adjusted return and drawdown is consistent with a risk-sensitive objective. Volatility is similar across objectives, so the SR gain comes from return shape and drawdown. This supports the sampling-based CE and Bellman-residual advantage as a valid risk-sensitive signal.

Markowitz under PPO (SR 1.43) trails recursive (SR 2.07); the per-step mean--variance reward is a different objective and may be harder in this setup. Equal-weight (1/N) is strong (SR 2.3--2.7 in the appendix); PPO underperforms it, so the benefit of recursive utility is \emph{relative} to the naive and Markowitz PPO baselines. High variance across splits and one trial per split imply the reported gains should be interpreted with caution; ablations (Appendix~\ref{app:ablation_full}) show sensitivity to $K$ and training window. Full results by algorithm and objective, and the ablation study, are in Appendix~\ref{app:results_full} and~\ref{app:ablation_full}.

% Main results: PPO by objective. Mean $\pm$ std over 10 splits; test.
\begin{table}[t]
\centering
\caption{PPO by objective (mean $\pm$ std over 10 splits). Full tables in Appendix~\ref{app:results_full},~\ref{app:ablation_full}.}
\label{tab:results_main}
\small
\resizebox{\textwidth}{!}{%
\begin{tabular}{lcccccc}
\toprule
\textbf{Obj.} & \textbf{SR} & \textbf{Sortino} & \textbf{Calmar} & \textbf{MDD (\%)} & \textbf{CR (\%)} & \textbf{Vol (\%)} \\
\midrule
Naive     & $1.22 \pm 1.07$ & $1.19 \pm 1.03$ & $4.10 \pm 5.28$ & $12.26 \pm 4.54$ & $-6.47 \pm 12.13$ & $23.42 \pm 4.06$ \\
Markowitz & $1.43 \pm 1.58$ & $1.74 \pm 2.01$ & $8.46 \pm 13.60$ & $13.74 \pm 9.73$ & $-3.68 \pm 16.27$ & $24.27 \pm 3.97$ \\
Recursive & $2.07 \pm 1.04$ & $2.11 \pm 1.09$ & $7.12 \pm 6.43$ & $10.38 \pm 4.26$ & $8.23 \pm 12.53$ & $25.05 \pm 7.89$ \\
\bottomrule
\end{tabular}%
}
\end{table}

\section{Conclusion}
\label{sec:conclusion}

We integrated recursive utility into actor-critic RL for portfolio allocation. The formulation is restricted to critic-based algorithms (PPO, A2C) because the certainty equivalent and the advantage both require a learned value function. On 10 chronological train/test splits of South Korean ETF returns, the recursive-utility agent improves on the naive (discounted) baseline in Sharpe ratio (2.07 vs 1.22), max drawdown (10.38\% vs 12.26\%), and cumulative return (8.23\% vs $-$6.47\%), consistent with a risk-sensitive objective. The appendices give derivations of the value function and advantage (AAE), experiment details (world model, environment, metrics), and additional results (hyperparameters, full performance tables, ablation study). Future work may add consumption-investment decisions and more efficient CE estimation; integrating forecasting models for alpha seeking or state-space models for state estimation and probabilistic forecasting could strengthen the world model. Broader deployment would require robustness under distribution shift and attention to regulatory and responsible-AI considerations.

% Bibliography (double-blind: no acknowledgments)
\bibliography{references}
\bibliographystyle{iclr2026_conference}

% Appendices (optional - only if needed and space permits)
\appendix
\section{Recursive Utility: Detailed Derivation}
\label{app:ez_derivation}

This appendix gives the recursive-utility aggregator, the fixed-point equation and operator $\mathcal{T}$, special cases, and closed-form approximations (Campbell--Viceira). Why the Bellman residual is used as the advantage is in Appendix~\ref{app:aae}. Theoretical properties of $\mathcal{T}$ (monotonicity, contraction) are in the main text; full derivations and proofs are in \cite{epstein1989substitution,schroder1999optimal,campbell2002strategic}.

\paragraph{Motivation.}
Recursive utility \cite{epstein1989substitution} separates risk aversion $\gamma$ from intertemporal elasticity $\psi$ (time-additive CRRA has $\gamma = 1/\psi$). The separation induces \emph{hedging demand} \cite{campbell2002strategic}; we test whether encoding it in the RL value target and advantage improves portfolio performance versus a discounted baseline.

\subsection{CES Aggregation}

CES utility: $U(x,y) = (\alpha x^\rho + (1-\alpha) y^\rho)^{1/\rho}$ with $\rho \in (-\infty, 1)$; elasticity of substitution $\sigma = 1/(1-\rho)$. Recursive utility applies this to current consumption and the certainty-equivalent of future utility:
\begin{equation}
U_t = \left( (1-\beta) C_t^\rho + \beta \left[\text{CE}_t(U_{t+1})\right]^\rho \right)^{1/\rho},
\label{eq:ces_ez}
\end{equation}
with $\rho = 1 - 1/\psi$. Limiting cases: $\rho \to 0$ (Cobb--Douglas), $\rho \to 1$ (linear), $\rho \to -\infty$ (Leontief); see \cite{epstein1989substitution}.

\subsection{Full Aggregator}

Combining CES time aggregation with CRRA risk aggregation:
\begin{equation}
V_t = \Big[ (1-\beta) C_t^{1-\frac{1}{\psi}} + \beta \Big(\mathbb{E}_t\big[V_{t+1}^{1-\gamma}\big]\Big)^{\frac{1-\frac{1}{\psi}}{1-\gamma}} \Big]^{\frac{1}{1-\frac{1}{\psi}}},
\label{eq:ez_full}
\end{equation}
with $\text{CE}_t[V_{t+1}] = (\mathbb{E}_t[V_{t+1}^{1-\gamma}])^{1/(1-\gamma)}$. Parameters: $\beta \in (0,1)$ (time preference), $\gamma > 0$ (risk aversion), $\psi > 0$ (IES); $\psi = 1/\gamma$ gives time-additive CRRA \cite{epstein1989substitution}.

\subsection{Fixed-Point Equation and Operator}

Wealth: $w_{t+1} = w_t + \log(1 + \boldsymbol{\alpha}_t^\top \mathbf{R}_{t+1})$ with $w_t = \log W_t$. Consumption $C_t = \kappa_t W_t$ is an accounting variable (main text); wealth dynamics do not depend on $\kappa_t$. Maximum recursive utility from state $s_t = (w_t, \boldsymbol{\alpha}_{t-1})$ satisfies
\begin{equation}
V(s_t) = \max_{\boldsymbol{\alpha}_t, \kappa_t} \Big\{ (1-\beta)(\kappa_t \exp(w_t))^{1-\frac{1}{\psi}} + \beta \Big(\mathbb{E}\big[V(s_{t+1})^{1-\gamma} \mid s_t, \boldsymbol{\alpha}_t, \kappa_t\big]\Big)^{\frac{1-\frac{1}{\psi}}{1-\gamma}} \Big\}^{\frac{1}{1-\frac{1}{\psi}}},
\label{eq:bellman_app}
\end{equation}
with $s_{t+1} = (w_{t+1}, \boldsymbol{\alpha}_t)$. Define the operator $\mathcal{T}$ by the right-hand side (same form with $s = (w, \boldsymbol{\alpha}_{\mathrm{prev}})$, $s' = (w', \boldsymbol{\alpha}')$):
\begin{equation}
(\mathcal{T}V)(s) = \max_{\boldsymbol{\alpha}, \kappa} \Big\{ (1-\beta)(\kappa \exp(w))^{1-\frac{1}{\psi}} + \beta \Big(\mathbb{E}\big[V(s')^{1-\gamma} \mid s, \boldsymbol{\alpha}, \kappa\big]\Big)^{\frac{1-\frac{1}{\psi}}{1-\gamma}} \Big\}^{\frac{1}{1-\frac{1}{\psi}}},
\label{eq:bellman_operator}
\end{equation}

\paragraph{Contraction.}
$\mathcal{T}$ is not a contraction in the supremum norm because of the CE nonlinearity. Under standard conditions (bounded value functions and consumption, $\beta < 1$), $\mathcal{T}$ is a contraction in the certainty-equivalent metric $d(V,W) = \sup_s |V(s)^{1-\gamma} - W(s)^{1-\gamma}|$ with modulus $\beta$. For precise conditions and proof we defer to \cite{epstein1989substitution,schroder1999optimal,sargent2025dynamic}.

\paragraph{Special cases.}
$\gamma = 1/\psi$: time-additive CRRA; $\mathcal{T}$ is linear and contracts in the supremum norm. $\psi = 1$: Cobb--Douglas ($\rho = 0$), multiplicative utility; limit $\rho \to 0$ needs separate treatment. $\gamma = 1$: CE becomes $\exp(\mathbb{E}[\log V])$; we assume $\gamma \neq 1$.

\subsection{Zero consumption}

With $C_t = 0$, $V_t = \big[ \beta (\mathbb{E}_t[V_{t+1}^\alpha])^{\rho/\alpha} \big]^{1/\rho}$ with $\alpha = 1-\gamma$, $\rho = 1-1/\psi$. Letting $Y_t = V_t^\alpha$ gives $Y_t = \tilde{\beta} \mathbb{E}_t[Y_{t+1}]$ with $\tilde{\beta} = \beta^{\alpha/\rho}$. Under homogeneity, $V_t$ can take the form $A(\mathbf{z}) W_t^\alpha$ for scale-invariant policies \cite{epstein1989substitution}.

\subsection{Two-period example}

Two periods $t=0,1$. Terminal value $V_1 = W_1^{1-1/\psi}$ (all wealth consumed). At $t=0$, $V_0$ is the CES of $C_0^{1-1/\psi}$ and $\text{CE}_0[V_1] = (\mathbb{E}_0[V_1^{1-\gamma}])^{1/(1-\gamma)}$. Substituting $V_1^{1-\gamma} = W_1^{(1-1/\psi)(1-\gamma)}$ and $W_1 = W_0(\alpha_0 R_1 + (1-\alpha_0)R_f)$ gives $V_0$ as a function of $\alpha_0$. For $C_0=0$ and $\psi=1$, maximizing $V_0$ is equivalent to maximizing $(\mathbb{E}_0[W_1^{1-\gamma}])^{1/(1-\gamma)}$; the FOC yields the Euler equation $\mathbb{E}_0[(\alpha_0 R_1 + (1-\alpha_0)R_f)^{-\gamma}(R_1 - R_f)] = 0$. Under lognormality, consumption growth and returns satisfy a log-linear Euler equation \cite{campbell2002strategic} with $\theta = (1-\gamma)/(1-1/\psi)$, which underlies the Campbell--Viceira approximate rules.

\subsection{Campbell--Viceira approximate rules}

With state $\mathbf{x}_t$ following a VAR and expected log excess returns $\mathbb{E}_t[\mathbf{r}_{t+1}^{\text{ex}}] = B \mathbf{x}_t$, log-linearization \cite{campbell2002strategic} yields
\begin{equation}
\boldsymbol{\alpha}_t^{\text{approx}} = \underbrace{\frac{1}{\gamma} \boldsymbol{\Sigma}^{-1} B \mathbf{x}_t}_{\text{myopic}} + \underbrace{\frac{1-\psi}{\psi} \boldsymbol{\Sigma}^{-1} \text{Cov}(\mathbf{r}_{t+1}, \mathbf{x}_{t+1}) \mathbf{A}_\alpha}_{\text{hedging}},
\label{eq:approx_portfolio_app}
\end{equation}
and $\log \kappa_t^{\text{approx}} = a_0 + \mathbf{a}_1^\top \mathbf{x}_t$. Myopic term depends on $\gamma$; hedging term vanishes when $\psi = 1$. We use these rules to initialize the actor (main text).

\section{Approximate Advantage Estimation}
\label{app:aae}

\subsection{Residual as advantage}

In standard RL, $A(s,a) = Q(s,a) - V(s)$ is an unbiased improvement signal. For recursive utility, the Bellman operator $\mathcal{T}$ (Appendix~\ref{app:ez_derivation}) is non-linear. The residual $(\mathcal{T}V_\phi)(s) - V_\phi(s)$ drives $V_\phi$ toward the fixed point (contraction) and is a valid policy-improvement signal; subtracting $V(s)$ keeps the policy gradient unbiased. In practice we use the one-step target \eqref{eq:ez_target} at the taken action, giving $\delta_t = \hat{T}_t^{EZ} - V_\phi(s_t)$ as in \eqref{eq:ez_advantage}. AAE below generalizes to multi-step with state-dependent weights; our experiments use AAE with $\lambda > 0$ (e.g.\ $\lambda = 0.95$); $\lambda = 0$ is the one-step residual.

\subsection{Why not GAE; TD error and multi-step}

Standard GAE \cite{schulman2015high} uses a linear Bellman equation and forms advantages as a linear combination of TD errors $\delta_t = r_t + \gamma V(s_{t+1}) - V(s_t)$. For recursive utility the recursion is non-linear (CE and outer power), so that construction is invalid; the $V = \mathbb{E}_\pi[Q]$ decomposition also fails. We define the action-fixed Bellman target $Q^{EZ}(s_t,a_t)$ as in Appendix~\ref{app:ez_derivation} and $A^{EZ}(s_t,a_t) = Q^{EZ}(s_t,a_t) - V(s_t)$. Approximating $Q^{EZ}$ by the one-step target $\hat{T}_t^{EZ}$ gives the TD error
\begin{equation}
\delta_t = \hat{T}_t^{EZ} - V_\phi(s_t),
\end{equation}
with $\hat{T}_t^{EZ}$ as in \eqref{eq:ez_target}. Multi-step recursive returns are not linear in these errors; they require nested non-linear aggregation. AAE approximates multi-step returns by a weighted sum of TD errors (next subsection).

\subsection{AAE formula}

AAE is a weighted sum of TD errors with state-dependent weights that correct for the non-linearity:
\begin{equation}
A_t^{AAE} = \sum_{k=0}^{\infty} (\beta \lambda)^k \delta_{t+k} \prod_{j=0}^{k-1} \omega_{t+j},
\label{eq:aae_formula}
\end{equation}
where $\omega_{t+j}$ (we use $\omega$ to avoid confusion with log wealth $w$) are
\begin{equation}
\omega_{t+j} = \beta \Big(V_\phi(s_{t+j+1})^{1-\gamma}\Big)^{\frac{1-\frac{1}{\psi}}{1-\gamma} - 1} V_\phi(s_{t+j+1})^{-\gamma} (1-\gamma).
\label{eq:weight_explicit}
\end{equation}
The factor $(\beta\lambda)^k$ plays the role of GAE's $(\gamma\lambda)^k$; the product $\prod_j \omega_{t+j}$ corrects for the aggregator. Implementation: $A_T = 0$ and $A_t = \delta_t + \beta \lambda \omega_t A_{t+1}$ for $t = T-1, \ldots, 0$. $\lambda \in [0,1]$ trades off bias and variance ($\lambda=0$: one-step; $\lambda=1$: full multi-step).

\subsection{Summary}

Our experiments use AAE \eqref{eq:aae_formula} with $\lambda = 0.95$. $\lambda = 0$ recovers the one-step residual \eqref{eq:ez_advantage}. The policy gradient update is $\mathbb{E}[\nabla_\theta \log \pi_\theta(a|s) \, A_t^{AAE}]$.

\section{Experiment Details}
\label{app:environment}

%=============================================================================
\subsection{World Model}
%=============================================================================

Portfolio allocation is modeled with observed asset returns $\mathbf{R}_t \in \mathbb{R}^n$ at each step $t$. Data are chronological: per-split \texttt{train.csv} and \texttt{test.csv}; \texttt{history.csv} for VarCov. No look-ahead. The agent observes log wealth $w_t = \log W_t$ and previous weights $\boldsymbol{\alpha}_{t-1}$; no auxiliary belief state. Wealth evolves by the self-financing rule:
\begin{equation}
W_{t+1} = W_t \big(1 + \boldsymbol{\alpha}_t^\top \mathbf{R}_{t+1}\big), \qquad w_{t+1} = w_t + \log\big(1 + \boldsymbol{\alpha}_t^\top \mathbf{R}_{t+1}\big).
\end{equation}
$W_0 = 1$, $w_0 = 0$. Transaction costs and market impact are not modeled; the agent is fully invested on the simplex $\Delta^n$ ($\alpha_{i,t} \geq 0$, $\sum_i \alpha_{i,t} = 1$).

%=============================================================================
\subsection{Environment}
%=============================================================================

\textbf{Data.} Daily ETF closing prices, South Korean market, from publicly available sources (pykrx and the KIS developer API); 110 ETFs (listed, sufficient history, exclusion of excessive missingness). Returns $R_{i,t} = (P_{i,t} - P_{i,t-1})/P_{i,t-1}$; preprocessing: handle missing data, winsorize outliers. Ten chronological train/test splits (train ratio 50\%--90\%); train strictly before test.

\textbf{State and action.} State $s_t = (w_t, \boldsymbol{\alpha}_{t-1})$; initial $\boldsymbol{\alpha}_0$ equal-weight. Actions: portfolio weights $\boldsymbol{\alpha}_t \in \Delta^n$. Policy outputs increments $\Delta \boldsymbol{\alpha}_t$; $\boldsymbol{\alpha}_t = \text{Proj}_{\Delta^n}(\boldsymbol{\alpha}_{t-1} + \Delta \boldsymbol{\alpha}_t)$ (add, clip to $[0,1]$, normalize).

\textbf{Reward and value target.} Naive: $r_t = \boldsymbol{\alpha}_t^\top \mathbf{R}_{t+1}$. Markowitz: $r_t = R_{\mathrm{port},t} - \lambda \sigma_t^2$. Recursive: no per-step reward in the target; learning uses $\hat{T}_t^{EZ}$ (Section~\ref{sec:method}); advantage from Bellman residual (Appendix~\ref{app:aae}). Discounted target $\hat{T}_t = r_t + \gamma V_\phi(s_{t+1})$; recursive target and CE sampling as in main text.

\textbf{Episode.} Fixed length (e.g., 252 steps); $w_{t+1} = w_t + \log(1 + \boldsymbol{\alpha}_t^\top \mathbf{R}_{t+1})$ from dataset returns. Training on \texttt{train.csv} only; one evaluation rollout on \texttt{test.csv} per split; metrics from test-period portfolio returns. Mean $\pm$ std over 10 splits, one trial per split \cite{liu2020finrl}. An optional three-layer mode (structural allocation, alpha tilt, risk controller) uses state $s_t = (\boldsymbol{\alpha}_t, c_t^\alpha)$ and actions $(\lambda, \kappa)$ \label{app:three_layer}; details in Appendix~\ref{app:implementation}.

%=============================================================================
\subsection{Metrics}
%=============================================================================

From test-period portfolio returns $\{r_t\}_{t=1}^{T}$ ($r_t = R_{\mathrm{port},t}$). Wealth: $W_0 = 1$, $W_t = \prod_{s=1}^{t}(1 + r_s)$. Risk-free rate $r_f = 0$ unless stated.

\textbf{Cumulative return (CR).} $\mathrm{CR} = (W_T - 1) \times 100\%$.

\textbf{Sharpe ratio (SR).} With $\bar{r} = \frac{1}{T}\sum_t r_t$, $\sigma_r = \sqrt{\frac{1}{T-1}\sum_t (r_t - \bar{r})^2}$: $\mathrm{SR} = \frac{\bar{r} - r_f}{\sigma_r} \sqrt{252}$ (annualized).

\textbf{Sortino.} Downside deviation $\sigma_{\mathrm{down}}$ over $t$ with $r_t - r_f < 0$; $\mathrm{Sortino} = \frac{\bar{r} - r_f}{\sigma_{\mathrm{down}}} \sqrt{252}$.

\textbf{Max drawdown (MDD).} $\mathrm{Peak}_t = \max_{0 \leq s \leq t} W_s$; $\mathrm{DD}_t = (\mathrm{Peak}_t - W_t)/\mathrm{Peak}_t$; $\mathrm{MDD} = \max_t \mathrm{DD}_t$; reported as MDD\% $= \mathrm{MDD} \times 100\%$.

\textbf{Volatility (Vol).} $\mathrm{Vol} = \sigma_r \sqrt{252}$; reported as Vol\%.

\textbf{Calmar.} Annualized return over MDD: $\bar{R}_{\mathrm{ann}} = (1 + W_T - 1)^{252/T} - 1$; $\mathrm{Calmar} = \bar{R}_{\mathrm{ann}}/\mathrm{MDD}$ ($\mathrm{MDD} > 0$).

\textbf{Information ratio (IR).} With benchmark $\{b_t\}$, active return $a_t = r_t - b_t$: $\mathrm{IR} = (\bar{a}/\sigma_a) \sqrt{252}$.

\section{Additional Results}
\label{app:implementation}

\subsection{Hyperparameters}

Representative values (see spec files for exact run configs): \textbf{Environment}: \texttt{max\_frame: 2520}, \texttt{episode\_length: 252}, \texttt{num\_assets: 110}, \texttt{reward: naive}. \textbf{PPO}: \texttt{gamma: 0.99}, \texttt{lam: 0.95}, \texttt{clip\_eps: 0.2}, \texttt{time\_horizon: 128}, \texttt{minibatch\_size: 64}, \texttt{training\_epoch: 4}, \texttt{val\_loss\_coef: 0.1}. \textbf{Recursive}: \texttt{beta: 0.99}, \texttt{gamma: 5.0} (risk aversion), \texttt{psi: 1.0}, \texttt{kappa\_init: 0.1}, \texttt{ce\_samples: 10}. \textbf{Network}: MLP, \texttt{hid\_layers: [128, 128]}, ReLU, Adam \texttt{lr: 0.02}.

\subsection{Performance}
\label{app:results_full}
% Full results: RL algorithm, objective. Mean +/- std over 10 splits.
% Fits portrait page; no landscape so the rest of the paper stays portrait.
\begin{table}[H]
\centering
\caption{Portfolio performance by algorithm and objective (mean $\pm$ std over 10 train/test splits). SR = Sharpe ratio, MDD = max drawdown (\%), CR = cumulative return (\%), Vol = volatility (annualized, \%). Eval on test.}
\label{tab:results_full}
\resizebox{\textwidth}{!}{%
\small
\begin{tabular}{@{}p{4.5em}lcccccc@{}}
\toprule
\multicolumn{1}{c}{\textbf{RL}} & \textbf{Objective} & \textbf{SR} & \textbf{Sortino} & \textbf{Calmar} & \textbf{MDD (\%)} & \textbf{CR (\%)} & \textbf{Vol (\%)} \\
\midrule
\multirow{2}{*}{Random}   & Naive     & $2.28 \pm 1.27$ & $2.19 \pm 1.32$ & $6.49 \pm 4.90$ & $6.68 \pm 1.42$ & $-0.12 \pm 7.35$ & $14.38 \pm 1.38$ \\
                                        & Markowitz & $2.66 \pm 1.17$ & $2.72 \pm 1.35$ & $9.51 \pm 9.55$ & $6.66 \pm 2.51$ & $1.70 \pm 6.32$ & $15.01 \pm 1.65$ \\
\addlinespace
\multirow{2}{*}{REINFORCE} & Naive     & $1.85 \pm 1.23$ & $1.94 \pm 1.36$ & $8.64 \pm 12.76$ & $9.70 \pm 3.77$ & $-0.44 \pm 10.17$ & $21.41 \pm 1.92$ \\
                                        & Markowitz & $1.37 \pm 1.03$ & $1.51 \pm 1.23$ & $5.28 \pm 7.05$ & $10.10 \pm 3.89$ & $-1.94 \pm 11.78$ & $23.06 \pm 2.78$ \\
\addlinespace
\multirow{3}{*}{A2C}      & Naive     & $1.78 \pm 4.05$ & $2.15 \pm 4.95$ & $11.74 \pm 29.56$ & $11.33 \pm 6.81$ & $-8.25 \pm 11.19$ & $21.18 \pm 9.85$ \\
                                        & Markowitz & $2.61 \pm 3.39$ & $3.02 \pm 4.84$ & $18.11 \pm 39.76$ & $12.62 \pm 11.10$ & $-2.83 \pm 14.68$ & $25.54 \pm 25.88$ \\
                                        & Recursive & $2.48 \pm 2.97$ & $2.85 \pm 3.52$ & $8.14 \pm 9.01$ & $8.87 \pm 5.68$ & $0.60 \pm 13.55$ & $20.20 \pm 10.20$ \\
\addlinespace
\multirow{3}{*}{PPO}       & Naive     & $1.22 \pm 1.07$ & $1.19 \pm 1.03$ & $4.10 \pm 5.28$ & $12.26 \pm 4.54$ & $-6.47 \pm 12.13$ & $23.42 \pm 4.06$ \\
                                         & Markowitz & $1.43 \pm 1.58$ & $1.74 \pm 2.01$ & $8.46 \pm 13.60$ & $13.74 \pm 9.73$ & $-3.68 \pm 16.27$ & $24.27 \pm 3.97$ \\
                                         & Recursive & $2.07 \pm 1.04$ & $2.11 \pm 1.09$ & $7.12 \pm 6.43$ & $10.38 \pm 4.26$ & $8.23 \pm 12.53$ & $25.05 \pm 7.89$ \\
\bottomrule
\end{tabular}%
}
\end{table}

\subsection{Ablation Study}
\label{app:ablation_full}
% Ablation: PPO recursive utility, CE K vs training window.
\begin{table}[H]
\centering
\caption{Ablation: PPO recursive utility. Rows = training window (episode\_length in days); columns = CE sampling $K$. Each cell: SR, Sortino, Calmar, MDD\%, CR\%, Vol\% (test period).}
\label{tab:ablation_full}
\resizebox{\textwidth}{!}{%
\small
\begin{tabular}{@{}lccccc@{}}
\toprule
\textbf{Training window} & \textbf{$K{=}1$} & \textbf{$K{=}2$} & \textbf{$K{=}5$} & \textbf{$K{=}10$} & \textbf{$K{=}20$} \\
\midrule
Whole year (window = 183) &
\begin{tabular}[t]{@{}l@{}}
SR: 4.00\\ Sortino: 4.26\\ Calmar: 18.78\\
MDD: 10.00\%\\ CR: 67.27\%\\ Vol: 27.41\%
\end{tabular} &
\begin{tabular}[t]{@{}l@{}}
SR: 1.83\\ Sortino: 1.78\\ Calmar: 2.86\\
MDD: 15.15\%\\ CR: 1.12\%\\ Vol: 20.95\%
\end{tabular} &
\begin{tabular}[t]{@{}l@{}}
SR: 0.08\\ Sortino: 0.07\\ Calmar: $-$0.01\\
MDD: 12.85\%\\ CR: $-$22.15\%\\ Vol: 17.44\%
\end{tabular} &
\begin{tabular}[t]{@{}l@{}}
SR: 1.52\\ Sortino: 1.37\\ Calmar: 3.35\\
MDD: 13.63\%\\ CR: 2.26\%\\ Vol: 27.13\%
\end{tabular} &
\begin{tabular}[t]{@{}l@{}}
SR: 1.53\\ Sortino: 1.53\\ Calmar: 3.11\\
MDD: 17.32\%\\ CR: 6.43\%\\ Vol: 31.46\%
\end{tabular} \\
\cmidrule(r){1-6}
Half year (window = 122) &
\begin{tabular}[t]{@{}l@{}}
SR: 2.01\\ Sortino: 1.75\\ Calmar: 5.56\\
MDD: 5.85\%\\ CR: $-$0.79\%\\ Vol: 14.56\%
\end{tabular} &
\begin{tabular}[t]{@{}l@{}}
SR: 1.90\\ Sortino: 1.84\\ Calmar: 6.22\\
MDD: 10.18\%\\ CR: 9.70\%\\ Vol: 27.89\%
\end{tabular} &
\begin{tabular}[t]{@{}l@{}}
SR: 0.82\\ Sortino: 0.87\\ Calmar: 1.63\\
MDD: 11.42\%\\ CR: $-$5.90\%\\ Vol: 24.52\%
\end{tabular} &
\begin{tabular}[t]{@{}l@{}}
SR: $-$1.25\\ Sortino: $-$1.32\\ Calmar: $-$1.50\\
MDD: 19.09\%\\ CR: $-$26.29\%\\ Vol: 24.55\%
\end{tabular} &
\begin{tabular}[t]{@{}l@{}}
SR: 0.47\\ Sortino: 0.35\\ Calmar: 0.78\\
MDD: 9.42\%\\ CR: $-$10.32\%\\ Vol: 18.59\%
\end{tabular} \\
\cmidrule(r){1-6}
Quarter (window = 61) &
\begin{tabular}[t]{@{}l@{}}
SR: 1.68\\ Sortino: 1.63\\ Calmar: 5.36\\
MDD: 7.49\%\\ CR: $-$3.57\%\\ Vol: 21.51\%
\end{tabular} &
\begin{tabular}[t]{@{}l@{}}
SR: 2.00\\ Sortino: 2.14\\ Calmar: 3.69\\
MDD: 16.47\%\\ CR: 6.21\%\\ Vol: 25.42\%
\end{tabular} &
\begin{tabular}[t]{@{}l@{}}
SR: 1.14\\ Sortino: 1.10\\ Calmar: 2.66\\
MDD: 8.04\%\\ CR: $-$12.81\%\\ Vol: 18.62\%
\end{tabular} &
\begin{tabular}[t]{@{}l@{}}
SR: 0.83\\ Sortino: 0.70\\ Calmar: 1.14\\
MDD: 17.35\%\\ CR: $-$13.65\%\\ Vol: 25.56\%
\end{tabular} &
\begin{tabular}[t]{@{}l@{}}
SR: 1.48\\ Sortino: 1.67\\ Calmar: 2.62\\
MDD: 18.32\%\\ CR: 0.16\%\\ Vol: 29.35\%
\end{tabular} \\
\cmidrule(r){1-6}
Month (window = 21) &
\begin{tabular}[t]{@{}l@{}}
SR: 2.02\\ Sortino: 2.10\\ Calmar: 2.83\\
MDD: 18.76\%\\ CR: 8.94\%\\ Vol: 22.38\%
\end{tabular} &
\begin{tabular}[t]{@{}l@{}}
SR: 0.89\\ Sortino: 0.81\\ Calmar: 1.09\\
MDD: 18.00\%\\ CR: $-$11.96\%\\ Vol: 23.06\%
\end{tabular} &
\begin{tabular}[t]{@{}l@{}}
SR: 0.25\\ Sortino: 0.26\\ Calmar: 0.17\\
MDD: 19.14\%\\ CR: $-$22.48\%\\ Vol: 24.01\%
\end{tabular} &
\begin{tabular}[t]{@{}l@{}}
SR: 0.75\\ Sortino: 0.74\\ Calmar: 0.83\\
MDD: 22.07\%\\ CR: $-$12.80\%\\ Vol: 27.69\%
\end{tabular} &
\begin{tabular}[t]{@{}l@{}}
SR: 2.19\\ Sortino: 2.22\\ Calmar: 5.09\\
MDD: 12.81\%\\ CR: 16.25\%\\ Vol: 24.29\%
\end{tabular} \\
\bottomrule
\end{tabular}%
}
\end{table}

\end{document}